\documentclass[aps,twocolumn]{revtex4}
\usepackage{amssymb}

\usepackage{amsmath}
\usepackage{graphicx}

\input{tcilatex}

\begin{document}

\title{Reduced Joule heating in nanowires}
\author{Fran\c{c}ois L\'{e}onard}
\email{fleonar@sandia.gov}
\date{\today }

\begin{abstract}
The temperature distribution in nanowires due to Joule heating is studied
analytically using a continuum model and a Green's function approach. We
show that the temperatures reached in nanowires can be much lower than that
predicted by bulk models of Joule heating, due to heat loss at the nanowire
surface that is important at nanoscopic dimensions, even when the thermal
conductivity of the environment is relatively low. In additition, we find
that the maximum temperature in the nanowire scales weakly with length, in
contrast to the bulk system. A simple criterion is presented to assess the
importance of these effects. The results have implications for the
experimental measurements of nanowire thermal properties, for thermoelectric
applications, and for controlling thermal effects in nanowire electronic
devices.
\end{abstract}

\maketitle

\address{Sandia National Laboratories, Livermore, California 94551}

The thermal properties of nanostructures such as nanowires, nanotubes,
graphene, and nanometric devices have attracted much attention recently
because of the unexplored physics of thermal transport at nanoscale
dimensions. In particular, the thermal properties of nanowires (NWs) have
been studied both for possible applications and for basic science, and an
important question is that of the thermal properties during current flow.
Indeed, Joule heating leads to reduced performance\cite{wang,endoh} and
breakdown \cite{westover,zhou} of nanowire electronic devices. It is also
utilized to probe NW electronic transport properties\cite{katzenmeyer}, and
to controllably functionalize nanowires\cite{park}. Joule heating is also
relevant to applications in thermoelectrics\cite{borca,hochbaum,abramson},
to current-heated magnetic nanowires\cite{you}, and is important in
assessing the reliability of field-emission devices based on NWs. In all of
these cases, the temperature distribution along the NW due to Joule heating
plays an important role, and so far analytical expressions for the full
nanowire geometry in contact with a thermal environment have not been
presented\cite{note}.

Here we provide such expressions and show that the nanometer size plays a
crucial role in determining the temperature distribution in Joule-heated
nanowires, due to heat flow from the surface that cannot be neglected due to
the high surface to volume ratio, even when the thermal conductivity of the
environment is relatively low. As a consequence, the temperatures reached
inside the nanowire are significantly lower than predicted by bulk models,
which are often used to model temperature distributions in NWs. The maximum
temperature also scales differently with the length of the NW, increasing
much more weakly with length. A criterion is proposed to assess the
importance of these effects, which can be used, for example, when evaluating
the need for more involved three-dimensional temperature simulation tools,
or when designing thermal measurement systems.

To be specific, we consider the system shown in Fig. 1a. There, a nanowire
of radius $R$ and length $L$ is thermally and electrically contacted by two
infinite planar contacts at $z=\pm L/2$ and held at temperature $T=0$. A
uniform electronic current $I$ flows through the NW, causing Joule heating.
In the steady-state, the temperature distribution between the two contacts
is given by%
\begin{equation}
\frac{\partial ^{2}T\left( r,z\right) }{\partial r^{2}}+\frac{1}{r}\frac{%
\partial T\left( r,z\right) }{\partial r}+\frac{\partial ^{2}T\left(
r,z\right) }{\partial z^{2}}=-p\theta \left( r-R\right)   \label{hc}
\end{equation}%
where $p=I^{2}\rho /(\pi R^{2})^{2}\kappa _{NW}$ with $\rho $ the NW
electrical resistivity and $\kappa _{NW}$ the NW thermal conductivity. The
key physics is found in the presence of the theta function on the right hand
side of this equation, and in the boundary conditions at the surface of the
NW. These boundary conditions can take different forms depending on the
thermal relation between the nanowire and its environment\cite{carslaw}.
Here we focus on the case of intimate contacts, but we expect the results to
be relevant to a broad range of situations. In addition, the case of
intimate contacts has the advantage of not requiring unknown interface
parameters, and is relevant to core/shell NWs with minor modifications to
account for the shell thickness.

For intimate thermal contacts, the temperature is continuous at the
interface $\left. T_{NW}\left( r,z\right) \right| _{r=R}=\left.
T_{env}\left( r,z\right) \right| _{r=R}$ , and because the environment
surrounding the NW has in general a different thermal conductivity $\kappa
_{env}$ the flux continuity at the surface gives a boundary condition%
\begin{equation}
\kappa _{NW}\left. \frac{\partial T_{NW}\left( r,z\right) }{\partial r}%
\right| _{r=R}=\kappa _{env}\left. \frac{\partial T_{env}\left( r,z\right) }{%
\partial r}\right| _{r=R}.  \label{bc}
\end{equation}%
When $\kappa _{env}=0$, the temperature gradients are entirely in the axial
direction and the problem reduces to the steady-state temperature
distribution in a bulk material 
\begin{equation}
\frac{d^{2}T_{bulk}(z)}{dz^{2}}=-p.
\end{equation}%
This equation is often referred to as that for ``one-dimensional'' heat
transport, and is often used to analyze the thermal properties of NWs. The
solution of the bulk equation for a NW between $z=-L/2$ and $z=L/2$ is%
\begin{equation}
T_{bulk}\left( z\right) =\frac{p}{2}\left( \frac{L^{2}}{4}-z^{2}\right) ,
\label{Tbulk}
\end{equation}%
with the maximum temperature%
\begin{equation}
T_{bulk}^{\max }=\frac{pL^{2}}{8}.
\end{equation}

\begin{figure}[tbp]
\centering \includegraphics[scale=0.6,clip=true]{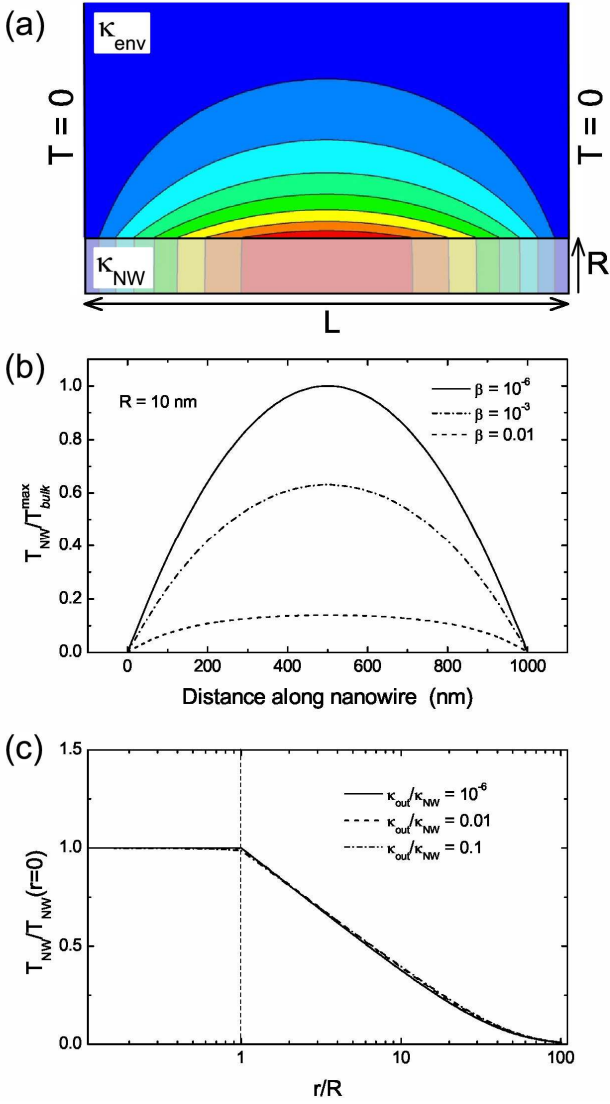}
\caption{Panel (a) shows the system under consideration: a nanowire of
length $L$ and radius $R$ between two plates held at $T=0$. The nanowire and
environment have thermal conductivities $\protect\kappa _{NW}$ and $\protect%
\kappa _{env}$. The color map shows a typical temperature profile with blue
corresponding to zero temperature and red to the largest temperature. Panel
(b) shows the temperature distribution along the length of a 10 nm radius
nanowire, plotted for three values of the ratio of thermal conductivities
between the environment and the nanowire. Panel (c) is the radial
temperature distribution normalized to the temperature at the origin, for
the same nanowire as in (b).}
\end{figure}

To obtain the solution for the NW, we first calculated the Green's function
for the differential equation in Eq. $\left( \ref{hc}\right) $ with the
boundary condition in $\left( \ref{bc}\right) $, and integrated over the
source term $p\theta \left( r-R\right) $. The solution is readily obtained as


\begin{eqnarray}
T(r,z)&=&\frac{4p}{L}\sum_{n=1,3,5,...}^{\infty }\sin \left( \frac{n\pi
\left( z+L/2\right) }{L}\right) \left( \frac{L}{n\pi }\right) ^{3}  \notag \\
&&\times \left[ 1- \frac{\beta I_{0}\left( \frac{n\pi r}{L}\right)
K_{1}\left( \frac{n\pi R}{L}\right) }{I_{1}\left( \frac{n\pi R}{L}\right)
K_{0}\left( \frac{n\pi R}{L}\right) +\beta I_{0}\left( \frac{n\pi R}{L}%
\right) K_{1}\left( \frac{n\pi R}{L}\right) }\right]
\end{eqnarray}


where $\beta =\kappa _{env}/\kappa _{NW}$, and $I_{\nu }$ and $K_{\nu }$ are
modified Bessel functions of order $\nu $. In the limit $\beta =0$ the
solution can be shown to reduce to that of the bulk case, Eq. $\left( \ref%
{Tbulk}\right) $.

Figure 1a shows the calculated temperature in the whole system for a general
case, showing the heat loss to the environment, while Fig. 1b shows the
temperature profile along the axis of a 10 nm radius NW of length 1 $\mu $m
for three values of the ratio of thermal conductivities $\beta $. While for
small values of $\beta $ the profile approaches that of the bulk solution,
for larger values of $\beta $ the maximum temperature reached in the
nanowire can be orders of magnitude lower. (This arises despite the fact
that the radial temperature gradients in the nanowire are still fairly
small, as shown in Fig. 1c.) Figure 2 shows the temperature reduction more
clearly by plotting the maximum temperature as a function of $\beta $ for
NWs of length 1 $\mu $m and of radii 1, 10, and 100 nm. The deviation from
the bulk solution depends strongly on the NW diameter, with the maximum
temperature dropping rapidly as $\kappa _{env}/\kappa _{NW}$ increases
beyond a value that is strongly diameter dependent.

As an example, for a 10 nm radius NW with the thermal conductivity of bulk
SiGe (10 W/mK) in air with thermal conductivity 0.045 W/mK, we have $\beta
=0.0045$ and the maximum temperature is almost a factor of five smaller than
the bulk solution would predict. The same NW surrounded by SiO$_{2}$ would
have $\beta \approx 0.1$ and the effect is even more striking. (While the
actual thermal boundary conditions for these systems might be different than
that given by Eq. (2), the same behavior is anticipated with $\beta $
replaced by the equivalent parameter describing the thermal properties of
the interfaces).

As expected, these effects become even more important for longer nanowires,
with a further reduction of the nanowire maximum temperature. However, in
addition to this qualitative expectation, there is also a change in the
scaling of the maximum temperature with length. Indeed, as Fig. 3 shows, the
bulk solution predicts a maximum temperature that scales with the length
squared; in contrast, for a nanowire with larger $\beta $, the scaling is
much different, and depends much more weakly on length.

\begin{figure}[tbp]
\centering \includegraphics[scale=0.8,clip=true]{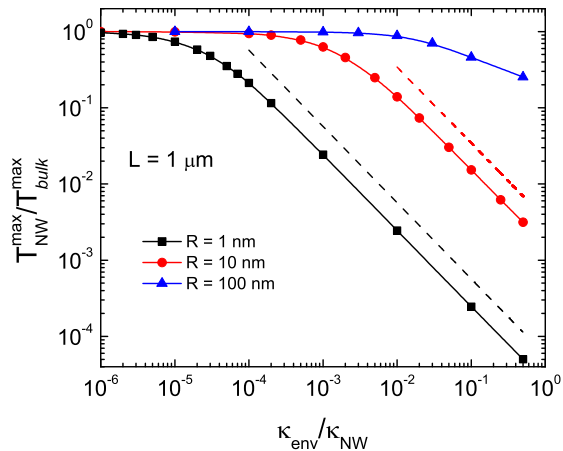}
\caption{Maximum temperature in Joule-heated nanowires as a function of the
ratio of the thermal conductivities of the environment and the nanowire. The
dashed lines are the predictions from Eq. (8).}
\end{figure}

To provide a simple expression detailing these effects we consider the first
mode in the above expansion, normalized by the first mode of the bulk
expression:%
\begin{equation}
\frac{T(r,z)}{T_{bulk}(z)}=1-\frac{\beta I_{0}\left( \frac{\pi r}{L}\right)
K_{1}\left( \frac{\pi R}{L}\right) }{I_{1}\left( \frac{\pi R}{L}\right)
K_{0}\left( \frac{\pi R}{L}\right) +\beta I_{0}\left( \frac{\pi R}{L}\right)
K_{1}\left( \frac{\pi R}{L}\right) }.
\end{equation}%
Expanding for small $\pi R/L$ we obtain for the temperature in the middle of
the nanowire%
\begin{equation}
\frac{T(0,z)}{T_{bulk}(z)}\approx \beta ^{-1}\left( \frac{\pi R}{L}\right)
^{2}\left| \ln \left( \frac{\pi R}{L}\right) \right| ,  \label{ratio2}
\end{equation}%
valid for $\beta \gg \left( \frac{\pi R}{L}\right) ^{2}\left| \ln \left( 
\frac{\pi R}{L}\right) \right| $. This expression is plotted in Fig. 2 for
the NWs of radii 1 and 10 nm; while the magnitude of this expression is
higher than the actual solution, it nevertheless represents the functional
dependence on $\beta $ very well, i.e. it decreases as $1/\beta $.
Furthermore, the dependence on $L$ is also well represented by this
expression, as the dashed line for the $R=10$ nm NW of Fig. 3 shows. By
including the $L^{2}$ dependence of the bulk solution, we get from Eq. $%
\left( \ref{ratio2}\right) $ the scaling $T_{NW}^{\max }\sim \left| \ln
\left( \frac{\pi R}{L}\right) \right| $ which shows the much weaker
dependence on $L$. From this analysis we deduce that the thermal
conductivity of the environment at which the bulk solution ceases to be
applicable is 
\begin{equation}
\kappa _{env}\succsim \kappa _{NW}\left( \frac{\pi R}{L}\right) ^{2}\left|
\ln \left( \frac{\pi R}{L}\right) \right|
\end{equation}%
indicating the importance of the aspect ratio $R/L$ and thus the nanoscale
dimensions. For a 10 nm radius NW of length 1 $\mu m$, deviations arise even
when $\kappa _{env}$ is only 0.1\% of $\kappa _{NW}$.

\begin{figure}[tbp]
\centering \includegraphics[scale=0.8,clip=true]{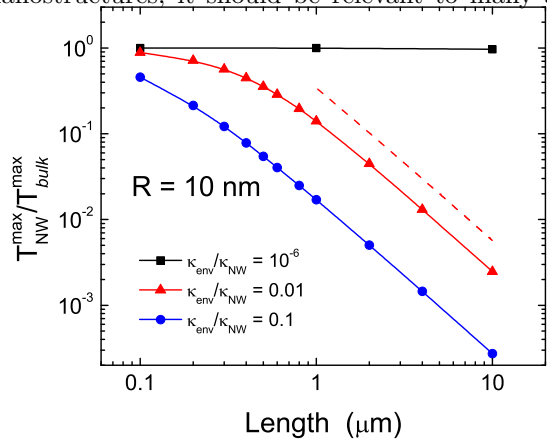}
\caption{Maximum temperature in Joule-heated nanowires as a function of the
length of the nanowire. The temperature is normalized by the maximum of the
bulk solution, which scales as $L^{2}$.}
\end{figure}

In summary, we presented a simple analytical expression for the temperature
distribution in Joule-heated nanowires. We find that the temperature reached
in nanowires during Joule heating can be significantly less than that
obtained from standard bulk models. This originates from heat loss at the
nanowire surface that cannot be ignored due to the high surface to volume
ratio, even when the thermal conductivity of the environment is relatively
low. The work presented here is straightforward, but given the ubiquitous
presence of Joule heating in nanostructures, it should be relevant to many
areas of nanoscale thermal transport, including thermoelectric devices based
on nanowire arrays, nanoscale transistors, and electrically driven
light-emitting nanodevices.

This project is supported by the Laboratory Directed Research and
Development program at Sandia National Laboratories, a multiprogram
laboratory operated by Sandia Corporation, a Lockheed Martin Company, for
the United States Department of Energy under Contract No. DEAC01-94-AL85000.

\end{document}